\documentclass[aps,prb,twocolumn,superscriptaddress,10pt,floatfix]{revtex4-2}
\usepackage[utf8]{inputenc}
\usepackage{xcolor}
\usepackage{amsmath}
\usepackage{graphicx}
\usepackage{booktabs}
\usepackage{float}

\begin{document}

\title{
Ultrafast Orbital-Selective Photodoping Melts Charge Order in Overdoped Bi-based Cuprates}

\author{Xinyi Jiang}\thanks{These authors contributed equally to this work.}
\affiliation{International Center for Quantum Materials, School of Physics, Peking University, Beijing 100871, China}

\author{Qizhi Li}\thanks{These authors contributed equally to this work.}
\affiliation{International Center for Quantum Materials, School of Physics, Peking University, Beijing 100871, China}

\author{Qingzheng Qiu}
\affiliation{International Center for Quantum Materials, School of Physics, Peking University, Beijing 100871, China}

\author{Li Yue}
\author{Junhan Huang}
\affiliation{International Center for Quantum Materials, School of Physics, Peking University, Beijing 100871, China}

\author{Yiwen Chen}
\affiliation{Beijing National Laboratory for Condensed Matter Physics, Institute of Physics, Chinese Academy of Sciences, Beijing 100190, China}

\author{Byungjune Lee}
\affiliation{Department of Physics, POSTECH, Pohang, Gyeongbuk, 37673, Republic of Korea}
\affiliation{
Max Planck POSTECH/Korea Research Initiative, Center for Complex Phase Materials, Pohang, Gyeongbuk, 37673, Republic of Korea}

\author{Hyeongi Choi}
\affiliation{PAL-XFEL, Pohang Accelerator Laboratory, POSTECH, Pohang, Gyeongbuk, 37673 Republic of Korea}

\author{Xingjiang Zhou}
\affiliation{Beijing National Laboratory for Condensed Matter Physics, Institute of Physics, Chinese Academy of Sciences, Beijing 100190, China}

\author{Tao Dong}
\author{Nanlin Wang}
\affiliation{International Center for Quantum Materials, School of Physics, Peking University, Beijing 100871, China}

\author{Hoyoung Jang}
\affiliation{PAL-XFEL, Pohang Accelerator Laboratory, POSTECH, Pohang, Gyeongbuk, 37673 Republic of Korea}
\affiliation{Photon Science Center, POSTECH, Pohang, Gyeongbuk, 37673 Republic of Korea}

\author{Yingying Peng}
\email{yingying.peng@pku.edu.cn}
\affiliation{International Center for Quantum Materials, School of Physics, Peking University, Beijing 100871, China}
\affiliation{Collaborative Innovation Center of Quantum Matter, Beijing 100871, China}

\date{\today}

\begin{abstract}

High-temperature superconductivity in cuprates remains one of the enduring puzzles of condensed matter physics, with charge order (CO) playing a central yet elusive role, particularly in the overdoped regime. Here, we employ time-resolved X-ray absorption
spectroscopy and resonant X-ray scattering at a free-electron laser to probe the transient electronic density of states and ultrafast CO dynamics in overdoped (Bi,Pb)$_{2.12}$Sr$_{1.88}$CuO$_{6+\delta}$.
We reveal a striking pump laser wavelength dependence - the 800\,nm light fails to suppress CO, whereas the 400\,nm light effectively melts it. This behavior originates from the fact that 400\,nm photons can promote electrons from the Zhang-Rice singlet band to the upper Hubbard band or apical oxygen states, while 800\,nm photons lack the energy to excite electrons across the charge-transfer gap. The CO recovery time ($\sim$3 ps) matches that of the underdoped cuprates, indicating universal electronic instability in the phase diagram. Additionally, melting overdoped CO requires an order-of-magnitude higher fluence highlighting the role of lattice interactions. Our findings demonstrate orbital-selective photodoping and provide a route to ultrafast control of emergent quantum phases in correlated materials.\\

\end{abstract}

\maketitle

High-temperature superconductivity (SC) has been a central focus of condensed matter physics since its discovery
\,\cite{zhou2021high, keimer2015quantum, paglione2010high}, and understanding the mechanism of SC and what controls superconducting temperature ($T_{\mathrm{c}}$) is a long-standing pursuit.  
Electronic correlations are pivotal in the formation of exotic phases intertwined with SC, and the electronic structure also significantly affects superconducting properties\,\cite{zhou2021high, wang2023correlating, ruan2016relationship, peli2017mottness}.
Cuprate superconductors achieved by doping the charge-transfer insulators exhibit the highest $T_{\mathrm{c}}$ at ambient pressure 
and has potential applications in energy and quantum technology. The critical temperature is considered related to the charge transfer gap (CT gap), and various experimental and theoretical studies have demonstrated their anticorrelation \,\cite{wang2023correlating, zou2022particle, ruan2016relationship, PhysRevB.72.014504,peli2017mottness,PhysRevResearch.2.043259,chakravarty2004explanation}. 
Furthermore, complex electronic correlations in the CuO$_2$ plane usually lead to exotic phases intertwined with SC, such as charge order (CO), stripe order, and pseudogap phase.

Notable competitions between CO and SC have been observed in underdoped cuprate superconductors \,\cite{chang2012direct,da2014ubiquitous,ghiringhelli2012long,lee2021spectroscopic,YBCOSW,YBCOhy}.
Furthermore, numerical calculations have found evidence for charge and/or spin density modulations as the ground state of the Hubbard model that competes with high-temperature SC\,\cite{zheng2017stripe,jiang2019superconductivity}. 
The overdoped regime is usually considered as the Fermi-liquid state, which can be described by BCS theory. However, the lower than expected superfluid density, the formation of Cooper pairs well above $T_{\mathrm{c}}$, and the ferromagnetic and nematic fluctuations provide evidence for unconventional SC behavior\,\cite{bovzovic2016dependence,uemura1993magnetic,he2021superconducting,kondo2015point,kurashima2018development,ayres2021incoherent}. 
Surprisingly, resonant X-ray scattering experiments have detected CO in the overdoped (Bi,Pb)$_{2.12}$Sr$_{1.88}$CuO$_{6}$ and heavily overdoped La$_{2-x}$Sr$_{x}$CuO$_{4}$\,\cite{ODCO,PhysRevLett.131.116002}. 
Unlike its underdoped counterpart, overdoped CO remains nearly unchanged up to room temperature, raising the
fundamental question whether overdoped CO shares the common origin as underdoped CO toward an ubiquitous instability. 
Elucidating the exotic states in the overdoped regime is crucial to understanding the evolution of $T_{\mathrm{c}}$ with hole doping and shedding light on the mechanism of high-$T_{\mathrm{c}}$ SC.

\begin{figure*}[htbp]
\centering
\includegraphics[width=0.85\linewidth]{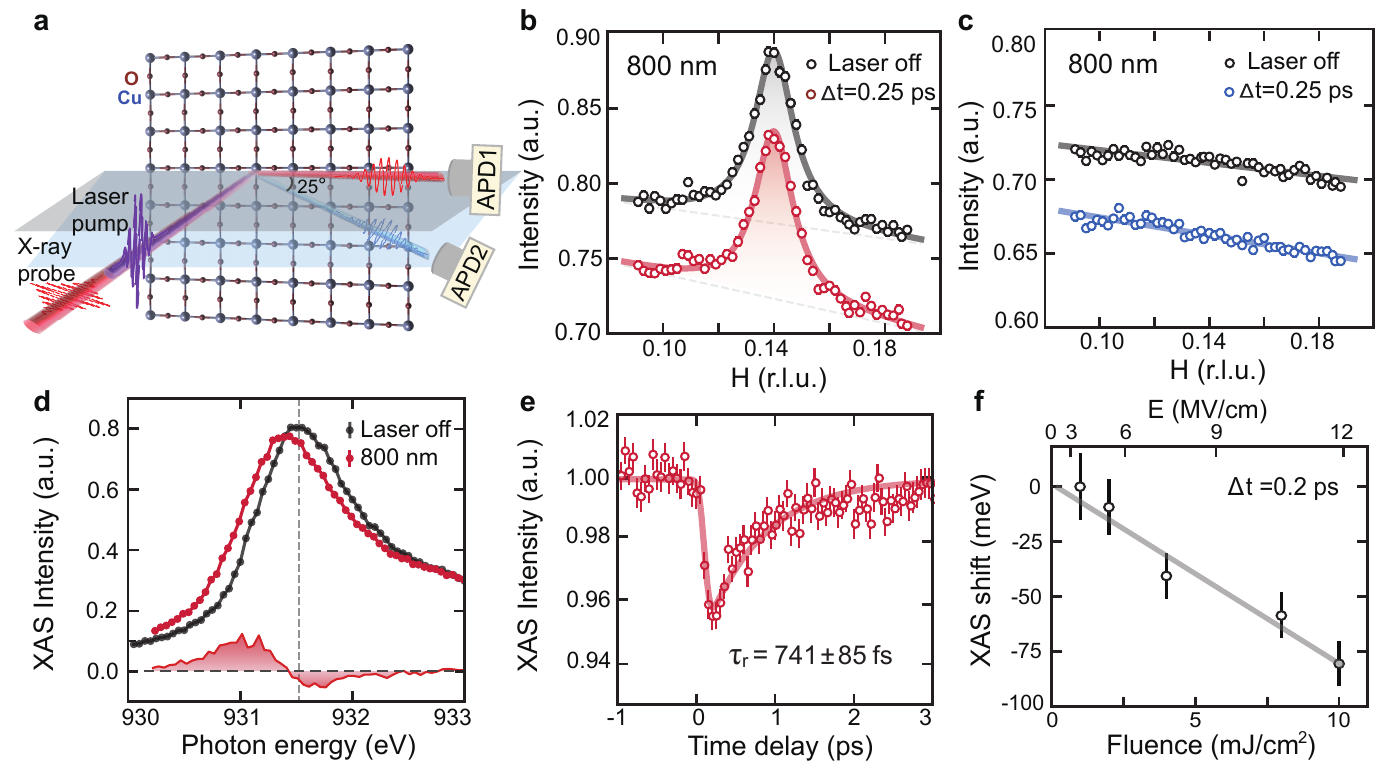} 
\caption{{\bf tr-REXS and tr-XAS in overdoped (Bi,Pb)$_{2.12}$Sr$_{1.88}$CuO$_{6}$.} 
{\textbf a} Schematic experimental configuration. 
Optical laser pulses perturb the sample, and X-ray pulses probe the CO peak. 
Two detectors are used: APD1 is parallel to the scattering plane to detect the CO scattering signal, and APD2 is positioned at a 25-degree deviation off the scattering plane to detect the fluorescence signal. 
{\textbf {b, c}} Transverse momentum scan in the H direction through the CO peak before (black curve) and after excitation ($\Delta$t = 0.25 ps, red and blue curve), induced by an 800\,nm pump with fluence of 5\,mJ/cm$^{2}$. The scattering signals {\textbf b} are fitted with the Lorentzian function and a linear background. The fluorescence signals {\textbf c} are fitted with a linear function. 
{\textbf d} XAS spectra in fluorescence-yield mode around Cu L$_{3}$-edge before and after pump excitation ($\Delta$t = 0.25\,ps) with fluence of 5\,mJ/c${\mathrm{m}}^{2}$. Three types of lines represent equilibrium (black), transient (red), and the corresponding difference spectrum. \textbf{e} Time dependence of XAS intensity at $\sim$931.5\,eV (dashed line in \textbf{d}), with 800\,nm (3\,mJ/c${\mathrm{m}}^{2}$) pumps. Intensities are normalized by the equilibrium absorption peak intensity. Error bars represent Poisson counting error. \textbf{f} XAS shift as a function of pump fluence. Experimental XAS shift values at the Cu $L_{3}$-edge are obtained by fitting the XAS peak with a pseudo-Voigt function on a linear background, resembling\,\cite{PhysRevX.12.011013}. The electric field is proportional to the square root of the laser fluence. The linear line is a guide to the eye. The error bars represent 1 standard deviation (s.d.) of the fit parameters. 
\label{fig1}}
\end{figure*}

An emerging approach to exploring the CO mechanism is to investigate its dynamics in the time domain using ultrafast techniques. Femtosecond time-resolved optical and photoemission spectroscopy can provide insight into the roles of electron-electron and electron-phonon interactions through time- and fluence-dependent behaviors \cite{PhysRevLett.102.066404,hellmann2010ultrafast,hellmann2012time,doi:10.1126/science.1160778,nakata2021robust}, yet these methods probe only the electronic structures rather than directly detecting CO. In recent years, time-resolved resonant X-ray scattering (tr-REXS) has enabled direct momentum-resolved measurements of the CO peak at Q$_{\rm CO}$, offering a more direct probe of CO dynamics. Previous tr-REXS studies on underdoped cuprates reveal diffusive
CO dynamics and the enhancement of CDW by suppressing SC with ultrafast laser pulses \cite{YBCOSW,YBCOhy,LBCOMM}.
Studying how the CO melts and reconstructs in the overdoped regime and comparing it to that of underdoped cuprates can clarify the interactions in the formation and stabilization of the ordered phases.

Here, we employ time-resolved X-ray absorption spectroscopy (tr-XAS) to track light-induced changes in the electronic density of states, alongside time-resolved resonant soft X-ray scattering (tr-REXS) to probe CO dynamics in overdoped (Bi,Pb)$_{2.12}$Sr$_{1.88}$CuO$_{6+\delta}$ (Bi2201, $T_{\mathrm{c}}$=11\,K, p$\sim$0.21). We observe a light-induced redistribution of the electronic density of states, consistent with an ultrafast renormalization of the on-site Coulomb interaction. Remarkably, the CO response shows a strong dependence on pump photon energy: 400 nm excitation melts the CO, while 800 nm light does not. This selective response reveals an orbital-dependent photodoping mechanism. We observe a CO recovery time of $\sim$ 3 ps, comparable to that in underdoped cuprates, pointing to a universal electronic instability across the phase diagram. Furthermore, an order-of-magnitude higher pump fluence is required to melt the CO in the overdoped regime, highlighting the important role of lattice coupling in stabilizing CO at high doping levels. Our results demonstrate that ultrafast optical pulses can selectively manipulate CO, offering a pathway to control quantum states in cuprates and other correlated materials.\\

\begin{figure}
\centering\includegraphics[width = \columnwidth]{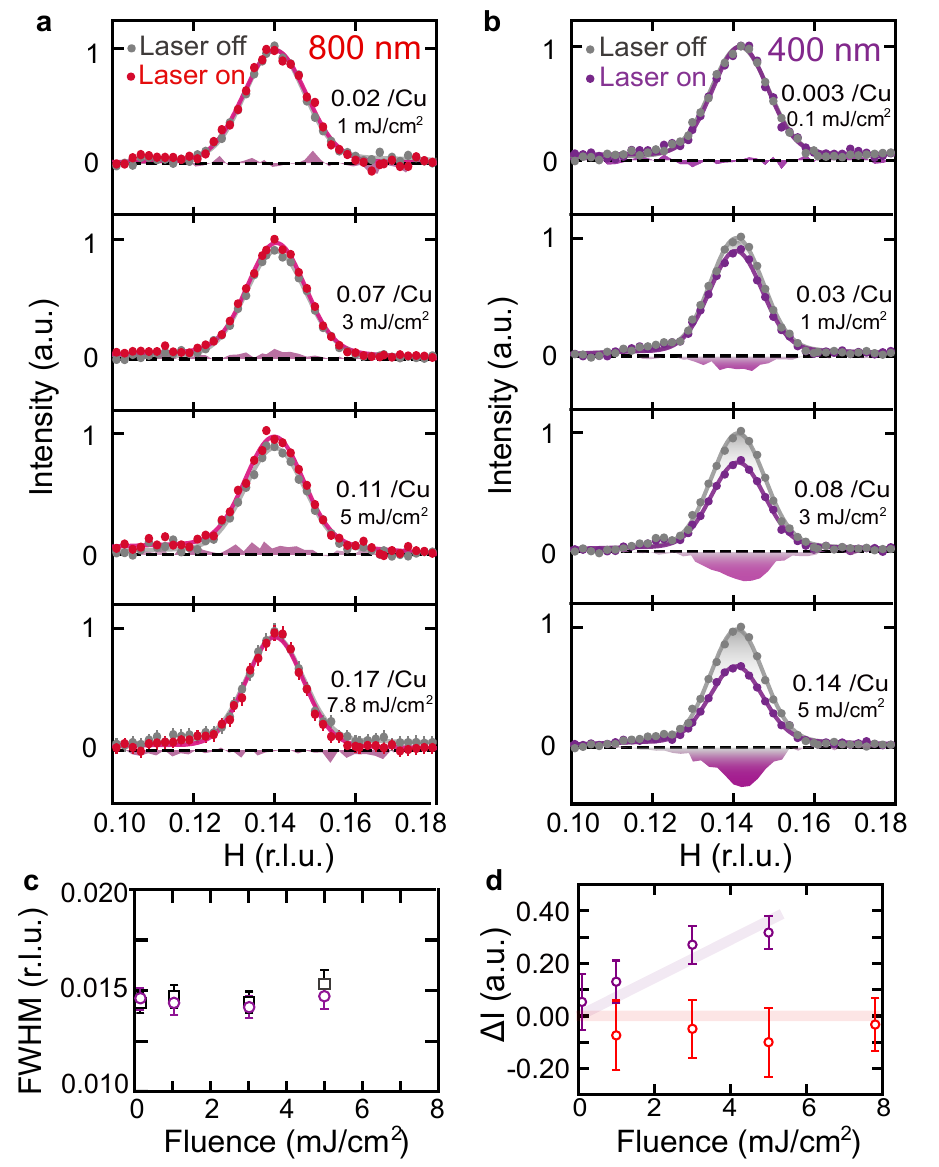} 
\caption{{\bf Pump-induced change in CO peak for the 800 and 400\,nm pumps.} \textbf{a, b} The evolution of the CO scattering peak divided by the fluorescent signal with different laser fluences for the 800\,nm {\textbf a} and 400\,nm {\textbf b} pumps at $\Delta$t=0.25\,ps at 80\,K. The purple shading indicates the difference between photo-induced states and equilibrium states. Solid lines are Lorentzian fits to the data with a linear background. 
\textbf{c} Fitted peak width (FWHM) of CO peaks before (black square) and after (purple circle) 400\,nm excitation as a function of fluence. The error bars represent 1 standard deviation (s.d.) of the fit parameters. 
\textbf{d} Changes in CO intensity induced by 800\,nm (red marker) and 400\,nm (purple marker) pumps. 
\label{fig2}}
\end{figure}

\begin{figure*}[htbp]
\centering
\includegraphics[width=0.85\linewidth]{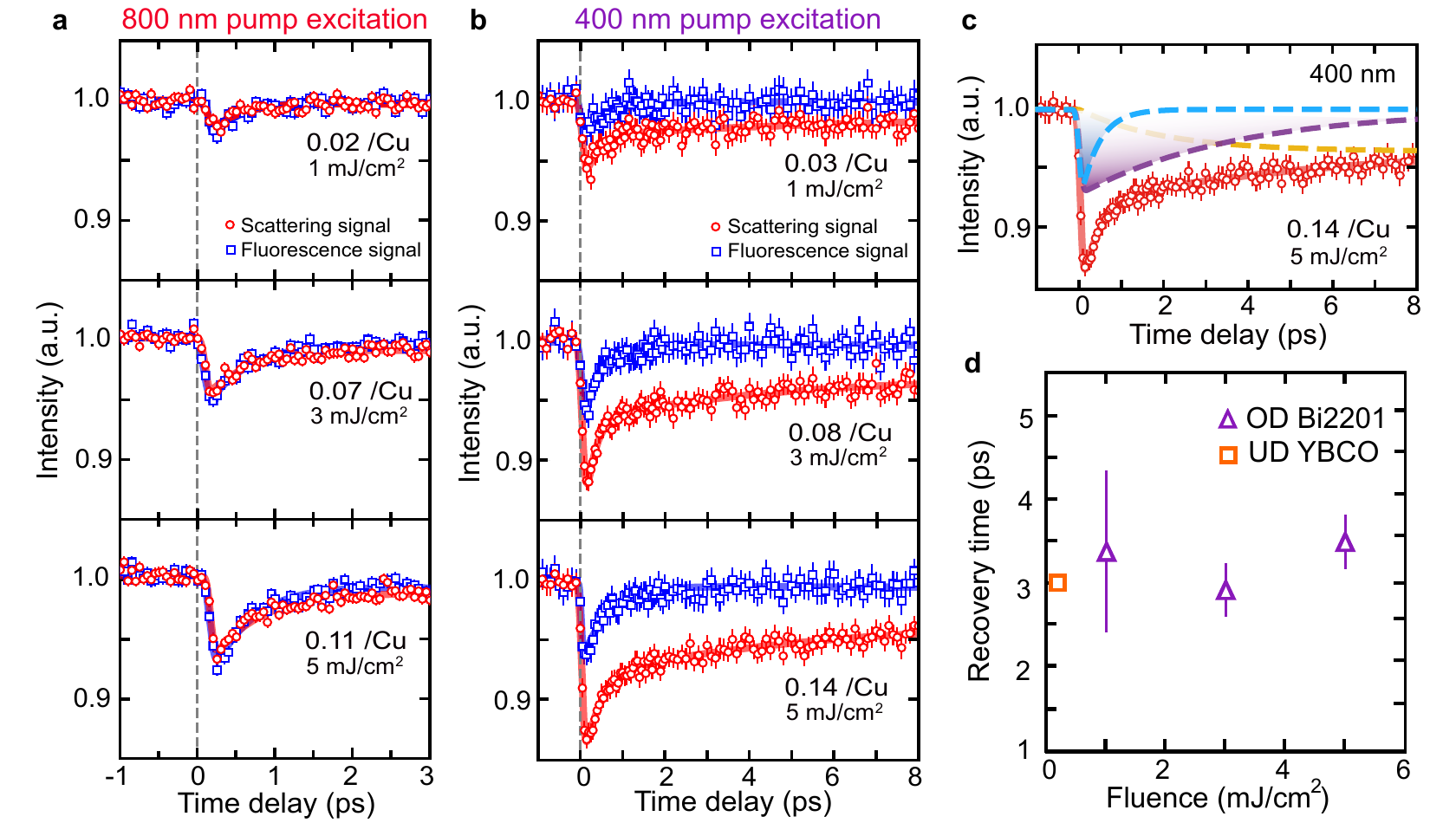}
\caption{{\bf Fluence dependence of the dynamics at the CO peak.} {\textbf {a, b}} Evolution of the CO peak intensity induced by 800\,nm {\textbf a} and 400\,nm {\textbf b} pumps with various fluences. 
The red and blue curves correspond to scattering and fluorescence signals, respectively. These intensities are normalized by the equilibrium scattering intensity. 
{\textbf c} X-ray scattering signal measuring the CO dynamics with a pump fluence of 5\,mJ/c${\mathrm{m}}^{2}$. The purple dashed line indicates a fit of the CO dynamics, and the blue dashed line indicates the electric dynamics. The orange dashed line represents the estimate of thermal heating effects. 
{\textbf d} Recovery time of CO dynamics as a function of fluence. The recovery time of CO on underdoped YBa$_2$Cu$_3$O$_{6.67}$ (YBCO) at 0.16\,mJ/c${\mathrm{m}}^{2}$ for an 800\,nm pump is included \cite{YBCOSW}. The error bars represent 1 standard deviation (s.d.) of the fit parameters. 
\label{fig3}}
\end{figure*}

\noindent
{\bf {RESULTS}}\\
\noindent
{\bf Light-induced change in the electronic density of states}

\noindent
To probe electronic modulations associated with the overdoped CO, we tune the incident X-ray energy to around Cu L$_3$-edge ($\sim$931.5\,eV), and two avalanche photodiode detectors (APD1 and APD2) are utilized to detect the scattering signal and the fluorescence signal, respectively (Fig.~\ref{fig1}\textbf{a}). We first use the 800\,nm laser to pump the overdoped CO as previous studies demonstrate that laser pulses at 800\,nm strongly perturb CO in underdoped cuprates. The CO peak is pronounced at the momentum transfer around the wave vector Q$_{\rm CO}$ = (0.14, 0, 3.5), consistent with previous quilibrium RIXS/REXS studies \cite{ODCO}. By applying the 800\,nm laser pump, the intensities in the scattering signal and the fluorescence signal are both uniformly suppressed (see Fig.~\ref{fig1}\textbf{b, c}). Meanwhile, the position and width of the CO peak do not change after photoexcitation. These photoinduced dynamics are in sharp contrast to the significant reduction in the intensity of the CO peak with unchanged background in underdoped cuprates\,\cite{YBCOSW,YBCOhy,LBCOMM}.

\begin{figure*}[htbp]
\centering
\includegraphics[width=0.85\linewidth]{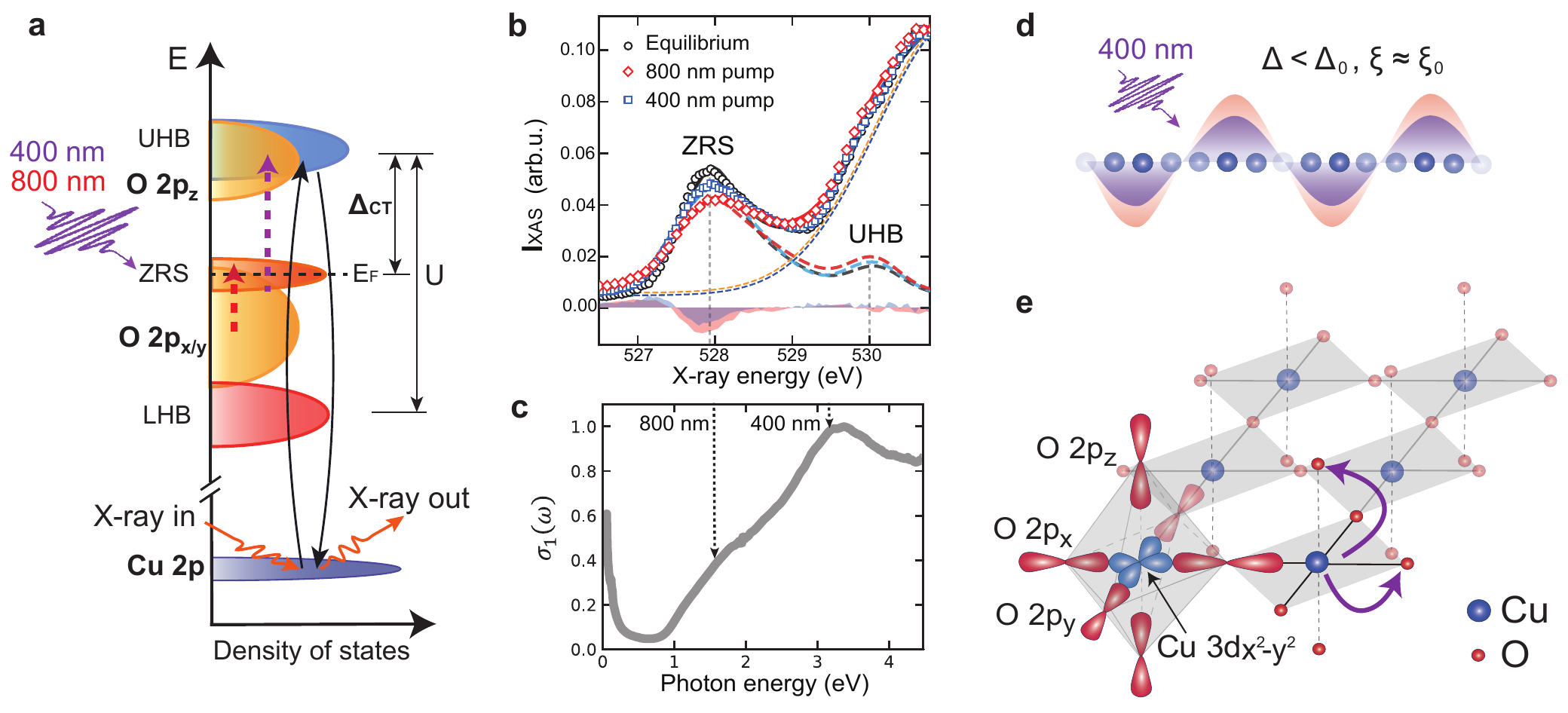}
\caption{{\bf 
Dynamical response of charge order induced by the 800\,nm and 400\,nm pumps.} \textbf{a} Schematic plot of tr-REXS process in the hole-doped cuprates. LHB, lower Hubbard band; ZRS, Zhang-Rice singlet peak; UHB, upper Hubbard band; $\Delta$$_{\rm CT}$, charge transfer gap. The purple and red arrows correspond to the 400 and 800\,nm pump respectively, while the orange and black arrows correspond to the X-ray probe process. 
\textbf{b} Time-resolved XAS spectra at O K absorption edges. The black curve indicates the equilibrium XAS spectrum, and the colored curves correspond to the transient XAS spectra at 0.25\,ps after the 800\,nm pump (red curve) and the 400\,nm pump (blue curve) with fluence of 5\,mJ/c${\mathrm{m}}^{2}$, corresponding to $\sim$0.11 absorbed photons/Cu for the 800\,nm pump and $\sim$0.14 absorbed photons/Cu for the 400\,nm pump. 
\textbf{c} Real part of the optical conductivity, $\sigma_{1}$($\omega$). 
\textbf{d} Schematic illustrating the dynamics of overdoped CO. The dashed grey line represents the equilibrium state of CO modulation. The solid purple line depicts the state excited by a 400\,nm pump pulse. $\Delta$ and $\xi$ denote CO amplitude and correlation length, respectively; $\Delta_0$ and $\xi_0$ are values at equilibrium.  Following photoexcitation, the CO amplitude is suppressed while the CO period and correlation length remain unchanged. \textbf{e} Sketch of the electron transfer process in the Cu-O octahedron of cuprates induced by the 400\,nm pump. 
\label{fig4}} 
\end{figure*}

To understand this intriguing non-equilibrium phenomenon in overdoped Bi2201, we trace the time-resolved XAS spectra at the Cu L$_{3}$-edge to probe the local electronic structure. 
As shown in Fig.~\ref{fig1}\textbf{d}, the absorption peak undergoes a significant redshift when comparing the equilibrium and transient spectra. The maximum energy of the transient XAS decreases by 80.3$\pm$10.1\,meV, leading to a uniform suppression of the scattering intensities as shown in Fig.~\ref{fig1}\textbf{b, c}. 
Figure~\ref{fig1}\textbf{e} displays the dynamics of the absorption peak intensity with an incident X-ray at 931.5\,eV (indicated by a dashed line in Fig.~\ref{fig1}\textbf{d}). The evolution of the intensity induced by an 800\ nm pump laser manifests itself as rapid decay after the pump (denoted ``quenching") and a subsequent recovery back to equilibrium. We use the following function to capture the relaxation:

\begin{equation}\label{eq:doubleExpFitting}
I(t) = I_{\rm 0} - \left[1+{\rm erf}(\frac{t}{\tau_{\rm q}})\right]A_{1}{\rm e}^{-\frac{t}{\tau_{\rm r}}}.
\end{equation}

Here, $I_{\rm 0}$ represents the initial intensity normalized by the equilibrium scattering peak intensity, and $\tau_{\rm r}$ represents the recovery time. 
The quench time is around 80$\pm$22\,fs limited by time resolution ($\sim$100\,fs), and the recovery time is 741$\pm$85\,fs, consistent with the holon-doublon recombination dynamics in cuprates\,\cite{PhysRevX.12.011013,PhysRevLett.111.016401,chargeMottinsulator,PhysRevLett.104.080401,PhysRevB.83.125102}.
Moreover, the XAS shift is
found to increase nearly linearly with pump fluence within our measured range  (Fig.~\ref{fig1}\textbf{f}). The transient change of XAS also affects the Bragg peak intensity as shown in Supplementary Fig.~1. 
Previous time-resolved XAS studies on underdoped LBCO observed similar phenomena, where the laser electric field ($\sim$10 MV/cm) is considered strong enough to reduce Coulomb repulsion U due to dynamical enhancement of dielectric screening\,\cite{PhysRevX.12.011013}. \\

\noindent
{\bf {CO dynamics dependence on pump laser wavelength and fluence}}\\
To reveal whether the pump laser suppresses the CO intensity, we divide the scattering signal by the fluorescence signal, as shown in Fig.~\ref{fig2}\textbf{a} (raw data are shown in Supplementary Fig.~3 and 4). We do not observe any CO suppression by the 800\,nm pump up to 7.8\,mJ/c${\mathrm{m}}^{2}$.  The corresponding photoexcited density per Cu is estimated to be 0.02, 0.07, 0.11, and 0.17 at fluence of 1, 3, 5 and 7.8\,mJ/c${\mathrm{m}}^{2}$, respectively (details are shown in Supplementary Fig.~5 and Note 3). 
In stark contrast, CO melts with increasing fluence when switching to the 400\,nm pump (Fig.~\ref{fig2}\textbf{b}). The CO peak intensity is unchanged at 0.1\,mJ/c${\mathrm{m}}^{2}$, which can easily melt the underdoped CO intensity\,\cite{YBCOSW,YBCOhy,LBCOMM}. This suggests that the overdoped CO is more robust to laser excitation than the underdoped CO, which agrees with its robust temperature behavior in the equilibrium state. We do not resolve any change in CO peak position and width (Fig.~\ref{fig2}\textbf{c}), indicating that CO period and the correlation length remain unchanged in the photo-excited transient state. These indicate nonthermal behaviors as opposed to the peak broadening with increasing temperature\,\cite{ODCO}, and different from the width broadening due to light-induced topological defects\,\cite{zong2019evidence}. In addition, the width of Bragg peak (0, 0, 2) is almost unchanged after pumping (Supplementary Fig.~2), which also suggests the non-thermal effect. 
Figure~\ref{fig2}\textbf{d} shows that the CO melting intensity increases with increasing fluence of 400\,nm pump. On the other hand, the intensity change remains nearly zero under the 800\,nm pump, even when the photons absorbed at each copper site are greater than those of the 400\,nm pump (compare Fig.~\ref{fig2}\textbf{a} and \textbf{b}).

We then trace the dynamics at the peak position of CO (Q$_{\rm CO}$) as the transient spectra are essential to distinguish ultrafast processes.
As shown in Fig.~\ref{fig3}\textbf{a}, the dynamics of the scattering and fluorescence signals collected by two APD detectors overlapped well with each other under the 800\,nm pump excitation. We can use Eq.~\eqref{eq:doubleExpFitting} to fit the dynamics. 
The intensity first undergoes a fast quench within a resolution-limited timescale ($\tau_{\rm q}$$\sim$0.1\,ps), followed by a subpicosecond recovery timescale $\tau_{\rm r}$ $\sim$0.5\,ps, which increases slightly with increasing fluence. These electronic dynamics are equal to the transient XAS change in Fig.~\ref{fig1}\textbf{e}. 
Moreover, we observed the same dynamics at a different momentum Q=0.08\,r.l.u., which is away from Q$_{\rm CO}$ (Supplementary Fig.~6). These indicate that the time evolution of the intensity induced by the 800\,nm pump arises from the photoinduced transient shift of XAS.

The CO dynamics at the 400\,nm pump is notably different from that of the 800\,nm pump. The scattering dynamics differs from the fluorescence dynamics, as shown in Fig.~\ref{fig3}\textbf{b}. 
The fluorescence dynamics again originates from the photoinduced XAS shift for the 400\,nm pump (Supplementary Fig.~1). However, the scattering signal under the 400\,nm pump shows a higher quench percentage, where the melting intensity increases with increasing fluence. These require additional recovery processes. 
We use the following function to fit the scattering dynamics induced by the 400\,nm pump, which includes fast quenching, two recovery processes, and a thermal process.

\begin{equation}\label{eq:ExpFitting} \begin{split}
I(t) = I_{\rm 0}-\left[1+{\rm erf}(\frac{t}{\tau_{\rm q}})\right]\left[A_{1}{\rm e}^{-\frac{t}{\tau_{\rm r1}}} + A_{2}{\rm e}^{-\frac{t}{\tau_{\rm r2}}}\right] \\
- A{'}(1-{\rm e}^{-\frac{t}{\tau '}}).
\end{split} \end{equation}

Here, $\tau_{\rm r1}$ represents the fast recovery, $\tau_{\rm r2}$ represents the slow recovery, and $\tau '$ denotes the lattice thermal timescale of $\sim$2\,ps\,\cite{thermal2ps,PhysRevLett.99.197001}. 
The decomposed processes are shown in Fig.~\ref{fig3}\textbf{c}. 
The fast recovery process is $\sim$0.5\,ps as the fluorescence signals due to photoinduced electronic dynamics. Of particular interest is the slow recovery process $\sim$ 3\,ps that can be assigned to CO melting dynamics,  
which coincides with the recovery time of CO modulations in underdoped YBa$_2$Cu$_3$O$_{6.67}$ and striped nickelate La$_{1.75}$Sr$_{0.25}$NiO$_{4}$ (Fig.~\ref{fig3}\textbf{d})\,\cite{YBCOhy,YBCOSW,lee2012phase}.  
These additional dynamics disappear at Q=0.08 r.l.u., when we move away from Q$_{\rm CO}$ (Supplementary Fig.~6). Therefore, the dynamics at the CO peak induced by the 400\,nm pump originates from the combination of electronic dynamics and CO reconstruction in overdoped Bi2201.\\

\noindent
{\bf {DISCUSSION}}

\noindent
The similar recovery timescales of CO in overdoped Bi2201 with underdoped YBCO imply a common electronic instability in the cuprate phase diagram. By melting CO in the overdoped regime, where CO is unexpectedly robust\,\cite{ODCO,PhysRevLett.131.116002}, the study hints at pathways to enhance transient SC, as seen in underdoped cuprates\,\cite{fausti2011light}. The excitation-energy dependence of the CO melting in overdoped Bi2201 provides fresh insight into understanding its microscopic origin. 
As illustrated schematically in  Fig.~\ref{fig4}\textbf{a}, the lower Hubbard band (LHB) and upper Hubbard band (UHB) arise from on-site Coulomb repulsion in the Cu 3d orbitals. Upon hole doping, the Zhang-Rice single (ZRS) band emerges because of hybridization between the hole on the planar Cu 3d orbital and the surrounding O 2p orbitals, forming the CT gap between the copper and oxygen states. Our O K-edge XAS spectra on Bi2201 reveal that the apical oxygen 2p$_z$ band overlaps with UHB, and the CT gap ($\Delta$$_{\rm CT}$) is $\sim$2\,eV (Fig.~\ref{fig4}\textbf{b}), consistent with previous optical and STM studies\,\cite{peli2017mottness,van2009doping,cai2016visualizing,giannetti2011revealing}. Therefore, 400\,nm ($\sim$3.1\,eV) pumping can be effectively absorbed, as shown by the optical
conductivity in Fig.~\ref{fig4}\textbf{c}. It can induce the planar charge transfer from the ZRS band to the UHB band, or promote electrons from the Cu-O
plane into apical oxygen states, as illustrated in Fig.~\ref{fig4}\textbf{e}. 
This photoinduced charge transfer changes the in-plane doping level, which can suppress the CO amplitude without changing the period and correlation length (Fig.~\ref{fig4}\textbf{d}). 
The out-of-plane electron transfer is difficult to recover, consistent with our observed prolonged CO relaxation behavior that persists for up to 500\,ps (Supplementary Fig.~7). Our observation aligns with theoretical predictions of orbitally selective photodoping in La$_{1.885}$Ba$_{0.115}$CuO$_{4}$, where photoexcitation redistributes holes from CuO$_{2}$ planes to charge reservoir layers, transiently increasing SC\,\cite{PhysRevB.104.174516}. 
In contrast, although the 800\,nm ($\sim$1.55\,eV) photons exceed the metallic
plasma edge ($\sim$1\,eV) (Fig.~\ref{fig4}\textbf{c}) and can excite electrons from the O 2p$_{x/y}$ band to the empty ZRS band, it is less absorbed and does not induce CO suppression. 
Our tr-XAS measurements at the O K-edge (Fig.~\ref{fig4}\textbf{b}) show that both 400\,nm and 800\,nm excitation suppress and broaden the ZRS peak due to photoinduced particle-hole excitations, consistent with prior observations in La$_{1.905}$Ba$_{0.095}$CuO$_4$\,\cite{PhysRevX.12.011013}. The weaker suppression at 400\,nm reflects its tendency to deplete the ZRS band, whereas 800\,nm excitation enhances occupancy by promoting electrons into it.
Notably, the 800\,nm laser already exceeds $\Delta$$_{\rm CT}$ ($\sim$1.24\,eV) in underdoped YBa$_{2}$Cu$_{3}$O$_{6+x}$ and $\Delta$$_{\rm CT}$ ($\sim$1.5\,eV) in La$_{2-x}$Ba$_{x}$CuO$_{4}$ \,\cite{lee2005electrodynamics,achkar2016orbital,lenarvcivc2014charge,golevz2019dynamics,okamoto2011photoinduced,YBCOSW,YBCOhy,LBCOMM}, thus it can effectively melt CO in underdoped cuprates.

In underdoped cuprates and stripe-ordered nickelates, CO melting is commonly observed at relatively low excitation fluences of 0.01-1\,mJ/c${\mathrm{m}}^{2}$, indicative of an underlying electronic instability\,\cite{YBCOSW,YBCOhy,LBCOMM,lee2012phase,chuang2013real}. The excitation density used here is an order of magnitude higher. 
We notice that similar high fluence is required to melt COs in BaNi$_{2}$As$_{2}$\,\cite{pokharel2022dynamics},  CsV$_{3}$Sb$_{5}$\,\cite{ning2024dynamical}, and TaSe$_{2}$ and NbSe$_{2}$\,\cite{nakata2021robust}, where electron-phonon coupling contributes significantly to CO formation. This resilience to laser fluence suggests that lattice interactions contribute importantly to the robustness of CO at high doping levels, consistent with the observed temperature stability of the CO\,\cite{ODCO,PhysRevLett.131.116002} and recent theoretical predictions\,\cite{liu2024charge}. To fully disentangle electronic and lattice contributions, future experiments utilizing phonon-resonant excitation, such as mid-infrared or THz-pump and X-ray-probe measurements, would be valuable. Such studies could reveal whether selectively driving specific phonon modes can suppress or enhance CO, as seen in the lattice-driven stripe dynamics of nickelates\,\cite{lee2017nonequilibrium}, thereby clarifying the interplay of charge, orbital and lattice degrees of freedom.

In summary, by combining time-resolved X-ray absorption spectroscopy and resonant X-ray scattering, we achieve momentum-resolved insights into non-equilibrium electronic structures in the overdoped cuprates. Our findings demonstrate that ultrafast, orbital-selective optical excitation provides a powerful approach to uncovering the intricate interplay between electronic and lattice degrees of freedom in CO dynamics. This approach offers a versatile framework for probing and controlling intertwined orders, such as charge, spin, and nematicity, across a broad class of quantum materials, including nickelates\,\cite{zhang2025spin,yi2024nature}, Kagome materials\,\cite{huang2025revealing}, and iron-based superconductors\,\cite{sprau2017discovery,bohmer2022nematicity,yi2017role}. By tuning the pump-photon energy to match interband transitions involving different orbital characters, this methodology enables selective access to ultrafast dynamics, paving the way toward light-driven control of quantum phases and the design of functional quantum materials.\\

\noindent
{\bf {METHODS}}\\
{\bf Sample characterization}

The single crystal of overdoped (Bi,Pb)$_{2.12}$Sr$_{1.88}$CuO$_{6}$ was grown using the floating zone method of the traveling solvent. The method of sampling growth and characterization has previously been reported\,\cite{lin2010high}. 
The single crystal used in this study has dimensions of about 2\,mm by 2\,mm by 0.2\,mm. The lattice was defined by using the pseudo-tetragonal unit cell with $a$ = $b$ = 3.83\,\AA\,and $c$ = 24.54\,\AA. 
The sample was cleaved and examined by X-ray diffraction measurements (XRD) before experiments to confirm its highly crystalline quality. 
Single crystal X-ray diffraction measurements were performed using the custom-designed X-ray instrument equipped with a Xenocs Genix3D Mo K$\alpha$ (17.48\,keV) X-ray sourse, which provides $\sim$\,2.5\,$\times$\,$10^7$ photons/sec in a beam spot size of 150\,$\mu$m at sample position\,\cite{PhysRevResearch.5.L012032}. 
The $T_{\mathrm{c}}$ value ($\sim$11\,K) of overdoped Bi2201 is defined at the onset temperature of the diamagnetism by magnetic measurement, performed with a 1\,Oe magnetic field applied along the c-axis using a SQUID magnetometer (MPMS XL1, Quantum Design), which shows a sharp transition width of approximately 1\,K.\\

\noindent
{\bf tr-RSXS measurements}

The tr-RSXS experiments were carried out at the SSS-RSXS endstation of PAL-XFEL \cite{jang2020trREXS}. The sample was mounted on a six-axis open-circle cryostat manipulator and measured at a base temperature of $\sim$80\,K using liquid nitrogen. The sample surface was perpendicular to the crystalline c axis and the horizontal scattering plane was parallel to the bc plane. X-ray pulses with pulse duration $\sim$ 80\,fs and 60 Hz repetition rate were used for the soft X-ray probe. The X-ray was linear horizontal polarized ($\pi$-polarization), and the photon energy was tuned to Cu L$_{3}$-edge (931.5\,eV). 
To trace CO, we fixed the scattering angle of the detector at 2$\theta$ = 163$^{\circ}$. For $Q_{\rm CO}$ ($H$) $\sim$ 0.14\,r.l.u., the $L$ value corresponds to 3.5\,r.l.u.. 

We utilized a Ti:sapphire laser to provide optical lasers at 1.55\,eV (800\,nm) and 3.1\,eV (400\,nm) with a pulse duration of $\sim$50\,fs and a repetition rate of 30\,Hz. The 800\,nm is converted to 400\,nm through the BBO crystal. The overall time resolution was $\sim$\,107\,fs, determined by measuring the pump-probe cross-correlation. The optical laser was nearly parallel to the incident X-ray beam, with an angle difference of less than $1^\circ$. 
The pump fluence ranged from 0.1 to 10\,mJ/c${\mathrm{m}}^{2}$ was
mainly used. The size of the X-ray spot in the sample position was $\sim$160\,(H)x240\,(V)\,$\mu{\mathrm{m}}^{2}$ (FWHM), while the diameter of the optical laser spot diameter was about 700\,$\mu$m for 800\,nm and 400\,$\mu$m for 400\,nm in FWHM. 
The X-ray repetition rate was twice that of the pump pulses, enabling comparison of diffraction signals before and after pump excitation. \\

\noindent
{\bf Optical measurements}

The optical conductivity data were obtained from the reflectivity spectra $R(\omega)$ at room temperature and under ambient conditions. The near-normal incidence ab-plane reflectivity $R(\omega)$ in the frequency range of 0.05 to 2.5 eV was measured using a Bruker Vertex 80V Fourier transform infrared (FTIR) spectrometer. In the higher frequency range of 2.5 to 5 eV, $R(\omega)$ was measured using a halogen and deuterium lamp as the light source and detected by a high-sensitivity NOVA spectrometer (Idea Optics, China). For the low-frequency extrapolation ($<0.05$\,eV), the Hagen-Rubens relation was used. 
The real part of the optical conductivity, $\sigma_1(\omega)$, was obtained from the measured reflectivity via Kramers-Kronig transformation for OD Bi2201. \\ 
\noindent
{\bf Data availability:} All data needed to evaluate the conclusions in the paper are present in the paper and/or the Supplementary Materials. The data are available from
the corresponding author upon request.


\vspace{2 ex}
\noindent
{\bf Acknowledgements:} The tr-RSXS experiments were performed using the RSXS instrument at PAL-XFEL (Proposal No. 2022-2nd-SSS-002) funded by the Ministry of Science and ICT of Korea. We acknowledge the valuable discussion with Yao Wang, Giacomo Ghiringhelli, Thomas P. Devereaux, Brian Moritz, Yuan Li, and Simon E. Wall. Y.Y.P. is grateful for financial support from the Ministry of Science and Technology of China (Grants No. 2024YFA1408702 and No. 2021YFA1401903) and the National Natural Science Foundation of China (Grant No. 12374143). X.J.Z's work is supported by the National Natural Science Foundation of China (Grant No. 12488201) and the National Key Research and Development Program of China (Grant No. 2021YFA1401800). N.L.W. acknowledges the support by the National Natural Science Foundation of China (Grant No. 12488201). B.L. and H.J. acknowledge the support by the National Research Foundation
grant funded by the Korea government (MSIT) (Grant No. RS-2022-NR068223).\\

\noindent
{\bf Author contributions:} 
X.J. and Y.P. proposed and designed the research. X.J., Q.L., Q.Q., B.L., H,C., H.J. and Y.P. carried out the tr-RSXS experiments. L.Y., J.H., T.D. and N.W. carried out the optical experiments. Y.C. and X.Z. provided the OD-Bi2201 single crystals. X.J. characterized the samples. X.J. and Y.P. analyzed the data and prepared the manuscript. All authors have read and approved the final version of the manuscript. \\

\noindent
{\bf Competing interests:} The authors declare that they have no competing interests. \\


\newpage

\setcounter{figure}{0}

\newcommand{\beginsupplement}{%
        \newpage ~\\
        \newpage
        \setcounter{table}{0}
        \renewcommand{\thetable}{S\arabic{table}}%
        \setcounter{figure}{0}
        \renewcommand{\thefigure}{S\arabic{figure}}%
     }
\renewcommand\thefigure{\arabic{figure}}
\renewcommand\thetable{S\arabic{table}}
\renewcommand\theequation{S\arabic{equation}}

\renewcommand{\figurename}{Supplementary Figure}

\title{Supplmentary Materials:\\
Ultrafast Orbital-Selective Photodoping Melts Charge Order in Overdoped Bi-based Cuprates}

\author{Xinyi Jiang}\thanks{These authors contributed equally to this work.}
\affiliation{International Center for Quantum Materials, School of Physics, Peking University, Beijing 100871, China}

\author{Qizhi Li}\thanks{These authors contributed equally to this work.}
\affiliation{International Center for Quantum Materials, School of Physics, Peking University, Beijing 100871, China}

\author{Qingzheng Qiu}
\affiliation{International Center for Quantum Materials, School of Physics, Peking University, Beijing 100871, China}

\author{Li Yue}
\author{Junhan Huang}
\affiliation{International Center for Quantum Materials, School of Physics, Peking University, Beijing 100871, China}

\author{Yiwen Chen}
\affiliation{Beijing National Laboratory for Condensed Matter Physics, Institute of Physics, Chinese Academy of Sciences, Beijing 100190, China}

\author{Byungjune Lee}
\affiliation{Department of Physics, POSTECH, Pohang, Gyeongbuk, 37673, Republic of Korea}
\affiliation{
Max Planck POSTECH/Korea Research Initiative, Center for Complex Phase Materials, Pohang, Gyeongbuk, 37673, Republic of Korea}

\author{Hyeongi Choi}
\affiliation{PAL-XFEL, Pohang Accelerator Laboratory, POSTECH, Pohang, Gyeongbuk, 37673 Republic of Korea}

\author{Xingjiang Zhou}
\affiliation{Beijing National Laboratory for Condensed Matter Physics, Institute of Physics, Chinese Academy of Sciences, Beijing 100190, China}

\author{Tao Dong}
\author{Nanlin Wang}
\affiliation{International Center for Quantum Materials, School of Physics, Peking University, Beijing 100871, China}

\author{Hoyoung Jang}
\affiliation{PAL-XFEL, Pohang Accelerator Laboratory, POSTECH, Pohang, Gyeongbuk, 37673 Republic of Korea}
\affiliation{Photon Science Center, POSTECH, Pohang, Gyeongbuk, 37673 Republic of Korea}

\author{Yingying Peng}
\email{yingying.peng@pku.edu.cn}
\affiliation{International Center for Quantum Materials, School of Physics, Peking University, Beijing 100871, China}
\affiliation{Collaborative Innovation Center of Quantum Matter, Beijing 100871, China}

\date{\today}

\maketitle

\noindent
{\bf SUPPLEMENTARY NOTES}\\

\noindent
{\bf Note 1. Time-resolved X-ray absorption spectra and Bragg peak dynamics}\\

We trace the time-resolved X-ray absorption spectra in overdoped (Bi,Pb)$_{2.12}$Sr$_{1.88}$CuO$_{6+\delta}$ triggered by intense near-infrared pump pulses (800 and 400\,nm), as shown in Supplementary Fig.~\ref{figS2}\textbf{a}. By comparing equilibrium and transient XAS spectra, we observe substantial photo-induced redshift at the Cu L$_{3}$ edge. The photoinduced shift of the UHB peak ($\sim$80.3$\pm$10.1\,meV) are similar at 400\,nm and 800\,nm lasers with the same fluence of 5\,mJ/c${\mathrm{m}}^{2}$. 
Supplementary Fig.~\ref{figS2}\textbf{b} shows the dynamics at the absorption peak (indicated by the vertical line) for the 800 and 400\,nm pumps, which is similar to the photoinduced XAS shift in La$_{1.905}$Ba$_{0.095}$CuO$_{4}$ \cite{PhysRevX.12.011013}. Using Eq.(1) in the main text to fit the dynamics, the recovery time is around 741$\pm$85\,fs with 800\,nm laser and 755$\pm$84\,fs with 400\,nm laser. These timescales are consistent with holon-doublon recombination dynamics in other cuprates\,\cite{PhysRevLett.111.016401,chargeMottinsulator,PhysRevLett.104.080401,PhysRevB.83.125102}.  
We also trace the profile of Bragg peak (0, 0, 2) before and after excitation at two different X-ray energies, as shown in Supplementary Fig.~\ref{TrBragg}. The widths are nearly unchanged, implying a nonthermal effect. The intensity of the Bragg peak collected at 932\,eV (on the right side of the resonant absorption peak) is suppressed (Supplementary Fig.~\ref{TrBragg}\textbf{a}), and the recovery time of the corresponding dynamics is $\sim$430\,fs (Supplementary Fig.~\ref{TrBragg}\textbf{c}). In contrast, the intensity of the Bragg peak collected at 931\,eV (on the left side of the resonant absorption peak) is enhanced (Supplementary Fig.~\ref{TrBragg}\textbf{b}), with the corresponding recovery time of $\sim$300\,fs (Supplementary Fig.~\ref{TrBragg}\textbf{d}). \\

\noindent
{\bf Note 2. Raw data of transient CO profile at different fluences }\\

We tune the X-ray energy to the Cu L$_3$-edge ($\sim$931.5\,eV) to probe the CO profile at the delay time of 0.25\,ps. The raw data for the 800\,nm pump at fluences of 1, 3, 5 and 7.8\,mJ/c${\mathrm{m}}^{2}$ are shown in Supplementary Fig.~\ref{figS4}. The scattering intensity is uniformly suppressed with increasing fluences (Supplementary Fig.~\ref{figS4}\textbf{a}), while the width of the CO peak remains unchanged after pump excitation. Alongside this, the fluorescence signal is also uniformly suppressed (Supplementary Fig.~\ref{figS4}\textbf{b}). 

The raw data for 400\,nm pump at different fluences are shown in Supplementary Fig.~\ref{figS5}. 
When using a 0.1\,mJ/c${\mathrm{m}}^{2}$ pump laser, the CO peak profile is almost unchanged after excitation, indicating that overdoped CO is more resistant to laser excitation than CO in underdoped cuprates and stripe-ordered nickelates\,\cite{YBCOSW,YBCOhy,LBCOMM,lee2012phase,chuang2013real}. In contrast to the 800\,nm pump, the suppression of the CO profile induced by the 400\,nm pump is not uniform - the suppression is more pronounced at the CO peak position than away from the CO peak. Yet, the fluorescence signal is uniformly suppressed at 400\,nm similar to 800\,nm (see Supplementary Fig.~\ref{figS5}\textbf{b}). \\ 

\noindent
{\bf Note 3. Calculation of laser photoexcited density}\\

In order to directly compare the effect of different pumps excitations on the sample, we derive the laser photoexcited density per copper site according to the laser fluence.
First of all, we calculate the number of absorbed photons, 
\begin{equation}\label{eq:excitationdensity}
N_{\rm abs} = \frac{(1-R)(1-e^{-1})F}{\lambda_{\rm p} \hbar \omega_{\rm p}}.
\end{equation}
where $N_{\rm abs}$ indicates the number of absorbed photons per unit volume, $R$ indicates the reflectivity of (Bi,Pb)$_{2.12}$Sr$_{1.88}$CuO$_{6+\delta}$, $F$ corresponds to the laser fluence, $\lambda_{\rm p}$ is the penetration depth of the pump laser, and $\hbar$$ \omega_{\rm p}$ is the energy of each pump photon. We obtain the values of $R$ and $\lambda_{\rm p}$ by optical measurements, as shown in Supplementary Fig.~\ref{Rk_Bi2201}\textbf{a}. 
For the 800\,nm pump, $R$$\sim$0.13; for the 400\,nm pump, $R$$\sim$0.17. 
Since laser penetration depth = laser wavelength/(2$\pi$$\cdot$$k$), we can calculate the penetration depth based on the absorption index ($k$) through optical measurements, as shown in Supplementary Fig.~\ref{Rk_Bi2201}\textbf{b}. For the 800\,nm pump, $k$$\sim$0.72, the penetration depth is about 180\,nm; for the 400\,nm pump, $k$$\sim$0.94, the penetration depth is about 70\,nm. Notably, the penetration depth of both lasers is greater than that of the X-rays at Cu L$_{3}$-edge. 
Then, since the volume of each Cu$^{2+}$ in overdoped Bi2201 is on the order of 10$^{-22}$cm$^{3}$, the excitation density per Cu$^{2+}$ is equal to $N_{\rm abs}$/$V_{\rm Cu^{2+}}$. 
Therefore, for the 800\,nm pump with laser fluences of 1, 3, 5 and 7.8\,mJ/c${\mathrm{m}}^{2}$, the corresponding laser photoexcited densities per copper site are 0.02, 0.07, 0.11 and 0.17 absorbed photons/Cu, respectively. For the 400\,nm pump with laser fluences of 0.1, 1, 3 and 5\,mJ/c${\mathrm{m}}^{2}$, the corresponding laser photoexcited densities per copper site are 0.003, 0.03, 0.08 and 0.14 absorbed photons/Cu, respectively.

In addition, the absorptivity of overdoped Bi2201 at 800\,nm and 400\,nm lasers is relatively similar (Supplementary Fig.~\ref{Rk_Bi2201}\textbf{b}), demonstrating that the different CO melting phenomena cannot be attributed to the difference in absorption. \\

\noindent
{\bf Note 4. Momentum dependence of scattering dynamics for the 800 and 400\,nm pumps
}\\

Supplementary Fig.~\ref{figS1}\textbf{a} and \textbf{b} demonstrate the scattering dynamics at and away from Q$_{\rm CO}$ (Q=0.14 and 0.08\,r.l.u., respectively) for the 800 and 400\,nm pumps. The dynamics at Q$_{\rm CO}$ and away from Q$_{\rm CO}$ are similar at the 800\,nm pump (Supplementary Fig.~\ref{figS1}\textbf{a}), which is due to photoinduced XAS shift. In contrast to the 800\,nm pump, the dynamics at two different momenta are different at the 400\,nm pump (Supplementary Fig.~\ref{figS1}\textbf{b}). 
These momentum-dependent dynamics imply the combination of electron and CO dynamics induced by the 400\,nm pump at Q$_{\rm CO}$.  \\

\noindent
{\bf Note 5. Long-time scan of CO dynamics induced by the 400\,nm pump}\\

We measure the change in intensity of the scattering peak over a time scale of up to 500\,ps, as shown in Supplementary Fig.~\ref{Bi2201longtimescan500ps}. At 9\,ps after 400\,nm excitation, the intensity recovers up to 98\% of the equilibrium value. Subsequently, up to 500\,ps, the intensity recovers rather slowly to nearly 0.99, but still not to the equilibrium value. This is consistent with the relaxation of the out-of-plane electron transfer required for a long time. \\




\noindent
{\bf SUPPLEMENTARY FIGURES}\\

\begin{figure*}[htbp]
\centering
\includegraphics[width=0.67\linewidth]{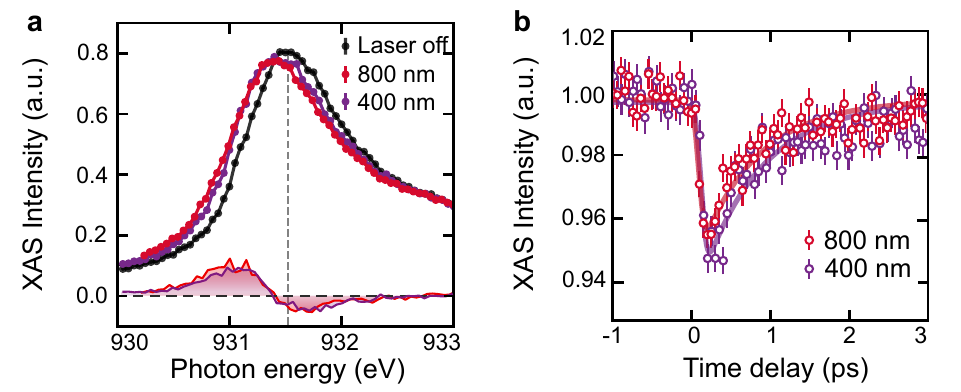}
\caption{{\bf Pump-induced X-ray absorption and electronic dynamics.} \textbf{a} XAS spectra in fluorescence-yield mode before and after pump excitation ($\Delta$t$\sim$0.25\,ps, fluence= 5\,mJ/c${\mathrm{m}}^{2}$, corresponding to $\sim$0.11 absorbed photons/Cu for the 800\,nm pump and $\sim$0.14 absorbed photons/Cu for the 400\,nm pump) arrival at Cu L$_{3}$-edge. 
Three types of lines represent equilibrium (black), transient (red for the 800\,nm pump and purple for the 400\,nm pump), and difference spectrum (shading). The equilibrium absorption peak is marked by dashed lines, which is found to shift after excitation. \textbf{b} The time evolution of the absorption peak at $\sim$931.5 eV, with excitation of 800\,nm (3\,mJ/c${\mathrm{m}}^{2}$, $\sim$0.07 absorbed photons/Cu) and 400\,nm (5\,mJ/c${\mathrm{m}}^{2}$, $\sim$0.14 absorbed photons/Cu), respectively. Intensities are normalized by the equilibrium absorption peak intensity. 
\label{figS2}}
\end{figure*}

\begin{figure*}[htbp]
\centering
\includegraphics[width=0.62\linewidth]{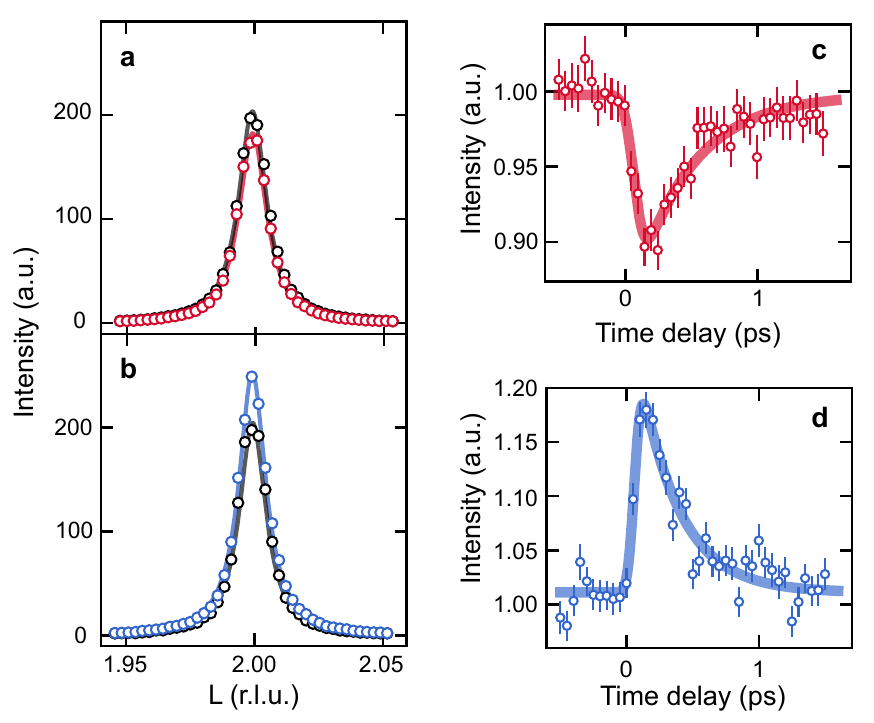}
\caption{{\bf Dynamics of Bragg peak (0, 0, 2) at different X-ray energies.} \textbf{a} Bragg peak (0, 0, 2) without pump and $\Delta$t=0.25\,ps after the 800\,pump with fluence of 5\,mJ/c${\mathrm{m}}^{2}$, with X-ray of 932\,eV (on the right side of the resonant absorption peak at 931.5\,eV). Solid lines are Lorentzian fits to the data. \textbf{b} Bragg peak (0, 0, 2) without pump and $\Delta$t=0.15\,ps after the 800\,pump with fluence of 5\,mJ/c${\mathrm{m}}^{2}$, with X-ray of 931\,eV (on the left side of the resonant absorption peak at 931.5\,eV). Solid lines are Lorentzian fits to the data. \textbf{c, d} Evolution of the Bragg peak with X-ray of 932\,eV \textbf{c} and 931\,eV \textbf{d}. Dynamics are fitted with Eq.(1) in the main text. 
\label{TrBragg}}
\end{figure*}

\begin{figure*}[htbp]
\centering
\includegraphics[width=0.75\linewidth]{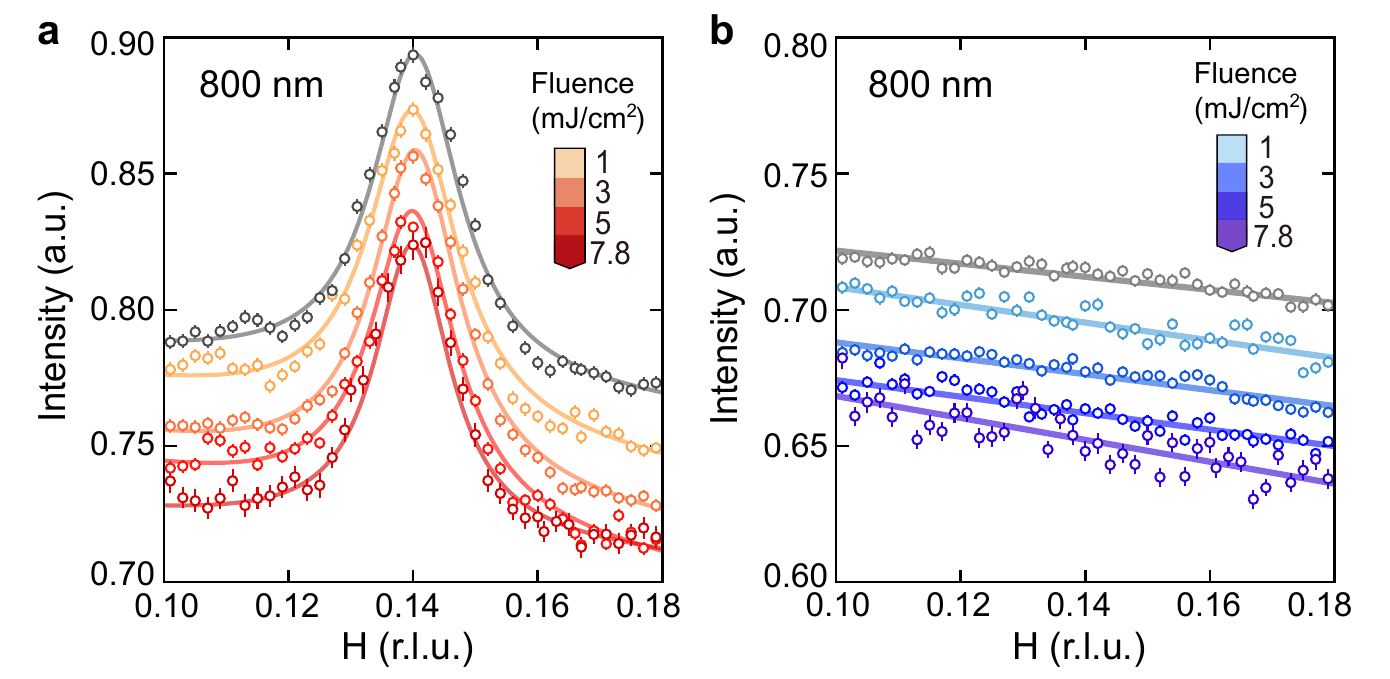}
\caption{{\bf Transient CO profile and fluorescence signal at 800\,nm pump.}
\textbf{a, b} The CO profile \textbf{a} and fluorescence signal \textbf{b} without pump (grey markers) and after pump excitation (0.25\,ps, 800\,nm) with fluence of 1, 3, 5 and 7.8\,mJ/c${\mathrm{m}}^{2}$ (color markers). Solid lines are fit using the Lorentzian function and a linear background. Error bars represent Poisson counting error. 
\label{figS4}}
\end{figure*}

\begin{figure*}[htbp]
\centering
\includegraphics[width=0.75\linewidth]{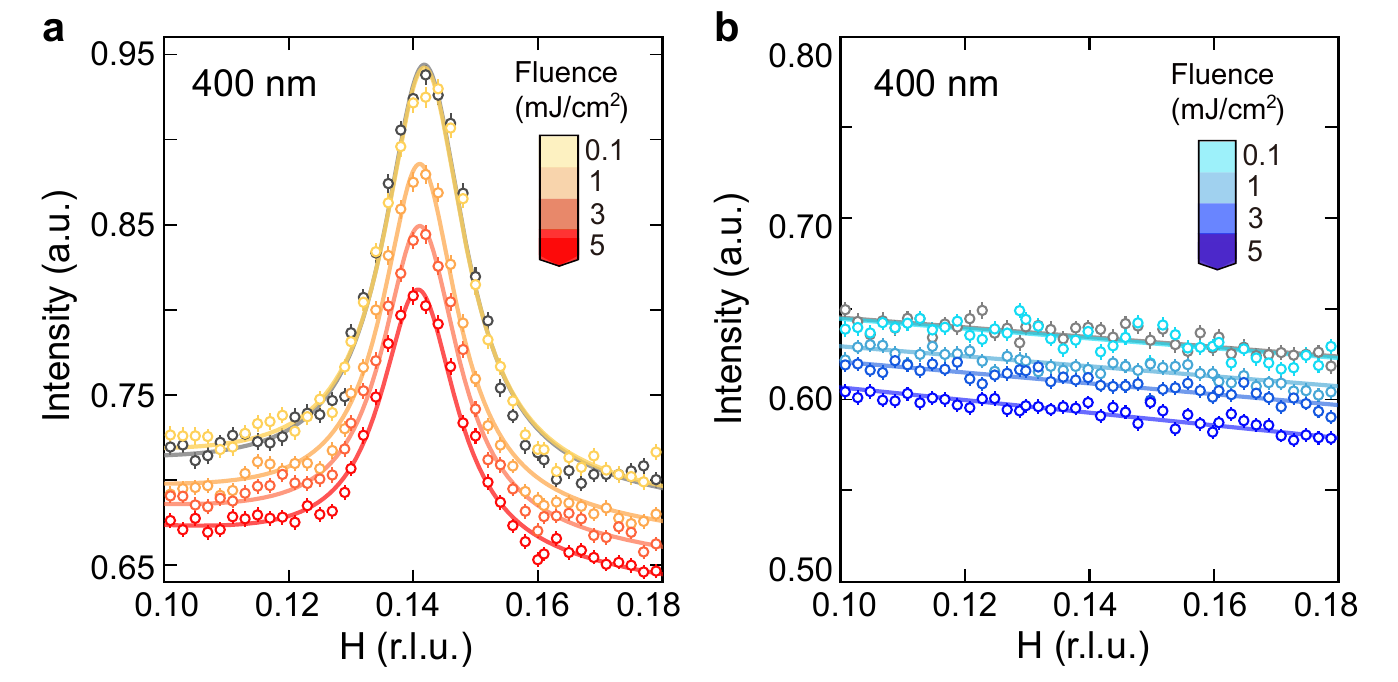}
\caption{{\bf Transient CO profile and fluorescence signal at 400\,nm pump.} \textbf{a, b} The CO profile \textbf{a} and fluorescence signal \textbf{b} without pump (grey markers) and after pump excitation (0.25\,ps, 400\,nm) with fluence of 0.1, 1, 3 and 5\,mJ/c${\mathrm{m}}^{2}$ (color markers). Solid lines are fit using the Lorentzian function and a linear background. 
\label{figS5}}
\end{figure*}

\begin{figure*}[htbp]
\centering
\includegraphics[width=0.8\linewidth]{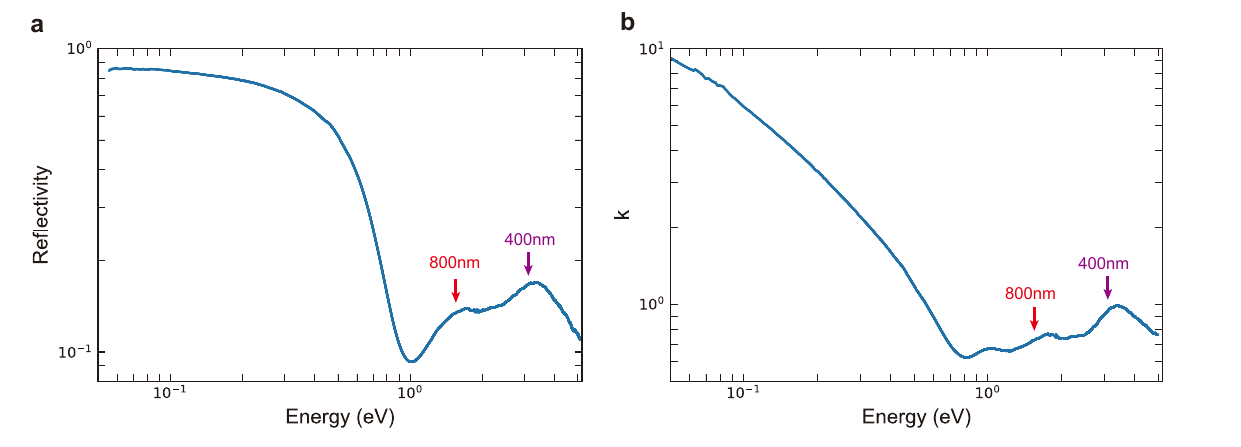}
\caption{{\bf Equilibrium optical measurements of overdoped (Bi,Pb)$_{2.12}$Sr$_{1.88}$CuO$_{6+\delta}$.}
\textbf{a, b} The refractivity \textbf{a} and absorption \textbf{b} as a function of laser wavelength. The red and purple arrows indicate 800\,nm and 400\,nm laser, respectively. 
\label{Rk_Bi2201}}
\end{figure*}

\clearpage

\begin{figure*}[htbp]
\centering
\includegraphics[width=0.7\linewidth]{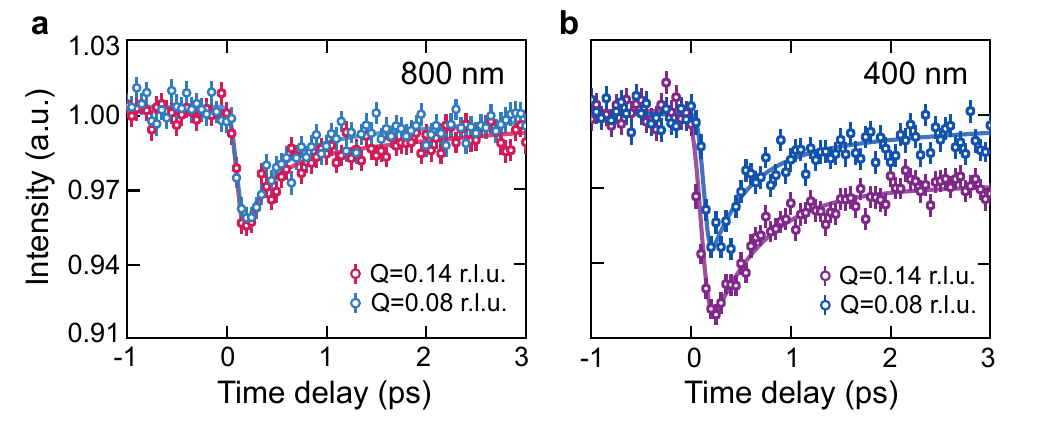}
\caption{{\bf Photoinduced dynamics at different momenta with excitation of 800\,nm and 400\,nm.} \textbf{a, b} Time evolution of scattering intensity at Q$_{\rm CO}$ and away from Q$_{\rm CO}$, with excitation of 800\,nm (3\,mJ/c${\mathrm{m}}^{2}$) in \textbf{a} and 400\,nm (5\,mJ/c${\mathrm{m}}^{2}$) in \textbf{b}. Intensities are normalized by the equilibrium scattering intensity. 
\label{figS1}}
\end{figure*}

\begin{figure*}[htbp]
\centering
\includegraphics[width=0.8\linewidth]{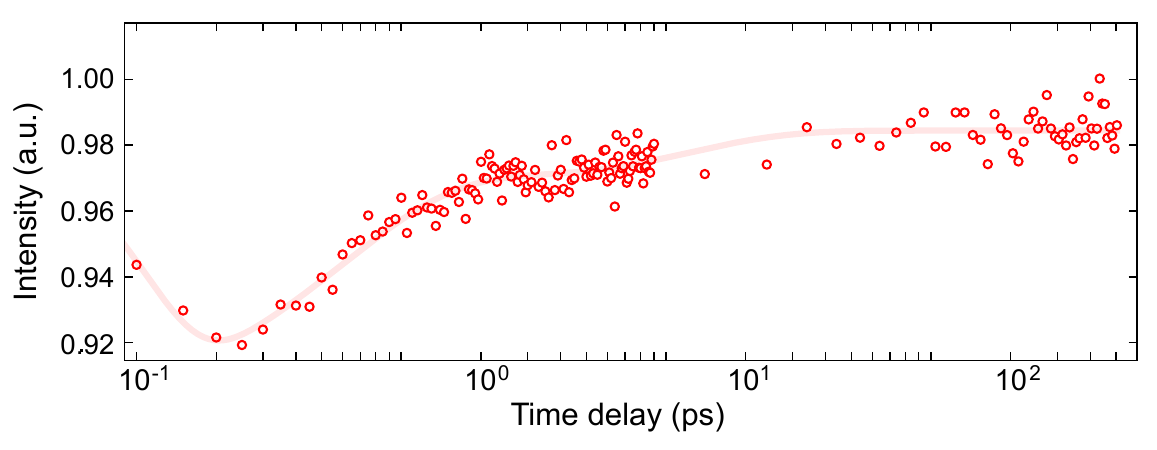}
\caption{ {\bf CO dynamics in a long-time scan.} 
Time evolution of CO peak intensity over a delay time of 500\,ps on a logarithmic scale, after the 400\,nm pump with fluence of 5\,mJ/c${\mathrm{m}}^{2}$. 
Intensities are normalized by the equilibrium peak intensity. 
\label{Bi2201longtimescan500ps}}
\end{figure*}


\end{document}